\documentstyle[aps]{revtex}

\begin{document}
\title{Recurrence quantification analysis as a tool for the characterization of
molecular dynamics simulations}
\draft
\author{Cesare Manetti and Marc-Antoine Ceruso}
\address{Department of Chemistry, University of Rome ``La Sapienza",\\
Piazzale Aldo\\
Moro, 5-00185 Rome, Italy}
\author{Alessandro Giuliani}
\address{Istituto Superiore di Sanita', TCE Lab, Rome 00161, Italy}
\author{Charles L. Webber, Jr.}
\address{Department of Physiology, Loyola University Medical Center, 2160\\
South First\\
Avenue, Maywood, IL 60153 USA}
\author{Joseph P. Zbilut}
\address{Department of Molecular Biophysics and Physiology, Rush\\
University, 1653 W.\\
Congress, Chicago, IL 60612 USA}
\date{\today}
\maketitle

\begin{abstract}
A molecular dynamics simulation of a Lennard-Jones fluid, and a trajectory
of the B1 immunoglobulin G-binding domain of streptococcal protein G
(B1-IgG) simulated in water are analyzed by recurrence quantification, which is noteworthy for its independence from stationarity constraints, as well as its ability to detect transients, and both linear and nonlinear state changes. The results demonstrate the sensitivity of the technique for the discrimination of phase sensitive dynamics. Physical interpretation of the recurrence
measures is also discussed.
\end{abstract}


\section{INTRODUCTION}

The quantitative analysis of molecular dynamics (MD) trajectories implies
the need for the individuation of salient phenomena embodied in time series
data: unique patterns in the dynamics require taxonomies. This need has
engendered, as a consequence, the use of classical multivariate data
analysis techniques such as principal components analysis (PCA) \cite
{garcia,amadei} and cluster analysis (CA) \cite{karpen}.

An important criterion for the choice of analysis method for MD trajectories
is the method's dependence on dynamical components of the data set, and its
relative independence from purely statistical characteristics. This
requirement is expressible in terms of ``phase information'' sensitivity,
and is broadly defined as having properties that are destroyed by random
shuffling of the series itself (shuffling sensitive information). From this
perspective, the usual statistical descriptors (e.g., mean, rms) do not
carry any phase information (shuffling resistant), while both PCA and CA
retain some information about the dynamics of the system (shuffling
sensitive).

In a previous paper \cite{giuliani}, we introduced recurrence quantification
analysis (RQA) as an alternative analysis technique to obtain phase
information about the energy landscape of simulated polypeptide systems.
This technique, proposed by Eckmann, Kamphorst and Ruelle as a purely
graphical tool and made quantitative by Zbilut and Webber \cite
{eckmann1,zbilut2,webber}, in contradistinction to PCA, gives a local view
of the studied series, based on single distance pairs and not
on the distribution of distances. This means that while autocorrelation
functions can only show general trends, RQA is particularly suited for the detection of
fast transients \cite{trulla}, and the consequent localization in time of
the salient features of the dynamics such as changes of state. Furthermore RQA's independence from stationarity constraints, ability to detect both linear and nonlinear dynamics \cite{webber,trulla,garcia}, and its ability to discriminate between signals and random noise \cite{zbilut1}, make RQA well-suited for a detailed characterization of MD trajectories (e.g.,
identification of microstates).

In this work, RQA was applied to the potential energy time series of
conformational space explored during MD simulations. The main goal was to
demonstrate the ability of RQA to discriminate the dynamics of a simple
system [a Lennard-Jones (LJ) fluid, which does not carry any phase
information, since it is a purely statistical, shuffling-resistant system]
from a complex system, such as the MD trajectory of B1-IgG simulated in
water (a protein that demonstrates shuffling-sensitive phase information,
due to the existence of structured paths between its microstates) \cite
{nienhaus}.

The testing strategy is straightforward: while in the case of an LJ fluid
the RQA measures must remain invariant after shuffling, they should change
significantly in the case of the protein. We will try to sketch a physical
characterization of RQA measures relative to MD.

\section{MATERIALS AND METHODS}

\subsection{LJ system MD simulations}

For the LJ simulation, we considered a system of 125 particles, enclosed in
a cube of side {\it L}, with periodic boundary conditions interacting
through a two-body potential of the LJ type:
\begin{equation}
V(r)=4[(\sigma /r)^{12}-(\sigma /r)^6]
\end{equation}
with the parameter, $\sigma =3.405$\AA ,\ corresponding to argon, so the
energies are expressed in units of $\epsilon $ $(\epsilon =119.8K)$ \cite
{verlet}. The simulations were performed at different temperatures and
varying {\it L} at different density. We used the same protocol for both
simulations: as a first step, the initial velocities were taken from a
Maxwellian distribution to perform 200 ps of simulation with only the last
100 ps being used for analysis. The potential energy time series was sampled
at 0.05 ps.

\subsection{Protein MD simulation}

All simulations of the protein were performed with the GROMACS simulation
package \cite{vdspoel}. A modification \cite{vanburen} of the GROMOS87 \cite
{vangunsteren1} force field was used with additional terms for aromatic
hydrogens \cite{vangunsteren2} and improved carbon-oxygen interaction
parameters \cite{vanburen}. SHAKE \cite{ryckaert} was used to constrain bond
lengths, allowing a time step of 2 fs.

The initial protein configuration was taken from the protein databank (1pga) 
\cite{gallagher}. The protein was immersed in a pre-equilibrated box of SPC
water \cite{berendsen1}, while 4 water molecules with the highest
electrostatic potential were replaced by sodium ions, resulting in an
electrically neutral cubic box ($a\simeq 4.1$ nm) containing 1790 water
molecules and 4 sodium counter ions for a total of 5936 atoms. Care was
taken that all crystallographic water molecules be conserved.

In order to prepare the solvated system for molecular dynamics, a three step
procedure was followed. Using a restraining harmonic potential, all heavy
atoms of the protein and the crystallographic water oxygens were constrained
to their initial positions while surrounding SPC water molecules were first
minimized and then submitted to 5 ps of constant volume MD at 300 K. The
resulting system was then minimized, without any constraints, before
starting constant temperature and constant volume MD. A nonbonded cutoff of
1.2 nm was used for both LJ and Coulomb potentials. The pair lists were
updated every ten steps. A constant temperature of 300 K was maintained by
coupling to an external bath \cite{berendsen2} using a coupling constant ($
\tau=0.002$) equal to the integration time step (2 fs). A total of 1.9 ns of
simulation were produced in this manner. The potential energy of the protein
was sampled every 0.1 ps.

\section{RECURRENCE QUANTIFICATION ANALYSIS (RQA)}

RQA was first introduced to the physical sciences by Eckmann, Kamphorst and
Ruelle in 1987 \cite{eckmann1} as a purely graphical technique. Five years
later Zbilut and Webber \cite{zbilut2} enhanced the technique by defining
five nonlinear descriptors of the recurrence plot that were found to be
diagnostically useful in the quantitative assessment of time series
structure in fields ranging from molecular dynamics to physiology \cite
{giuliani,mestivier,faure}. This technique has been demonstrated to be
particularly useful in quantifying transient behavior far from equilibrium
in relatively short time series \cite{trulla}. This feature is particularly
important in the detection of unique patterns in complex data sets \cite
{zbilut3}.

RQA is based on the computation of a Euclidean distance matrix between the
rows (epoch) of an embedded matrix of the scalar time series (in this case the MD simulations of potential energy) at a fixed lag (time delay method), as originally suggested by Ruelle and recorded by Packard \cite{packard}. Takens' \cite{takens} theorem states a mathematical relation between the embedding dimension of a suitable scalar observable and the real dimension of the attractor of the corresponding dynamical system $(d)$: $n = 2d + 1$, where $n$ is the
minimum embedding dimension to assure the reconstructability of the
underlying attractor. This relation holds true for deterministic, noiseless,
systems. In the presence of noise from whatever source (e.g., system, quantization), however, it has been shown that higher embeddings are necessary \cite{ding}. From this it is obvious that the embedding be sufficiently high so as to capture all the essential dynamics: the error
of choosing a too high an $n$ is not as great as choosing too low a value of $n$.
This is particularly true in MD simulations, where it is crucial to have sufficient
dimensionality to estimate higher order moments giving information on
correlations between the movements of two or more atoms in a protein \cite{clarage}. In the present case, 10 was dictated by our previous work on a tetrapeptide dynamics \cite{giuliani}.

Thus, the sequence of vectors (${\bf x}_i$), embedded
in $\Re ^n$, define a function on an $n\times n$ array according to the rule: darken the ($i,j$)th element of the array if ${\bf x}_j\in B({\bf x}_i,r)$, where $B(
{\bf x}_i,r)$ is the ball or radius $r$, centered at ${\bf x}_i$ (see \cite
{giuliani,webber} for details). The features of the distance function make
the plot symmetric $D_{i,j}=D_{j,i}$ with a darkened main diagonal
corresponding to the identity line ($D_{i,j}=0;$ $j=i$). The darkened points
individuate the recurrences (recurrent points) of the dynamical process and
the plot can be considered as a global picture of the autocorrelation
structure of the system. Consequently, a recurrence plot visualizes the
distance matrix, which, in turn, represents the autocorrelation present in
the series at all possible time scales. In fact, it is important to note
that the distance is computed for all the possible pairs of epochs, the
elements near the principal diagonal of the plot corresponding to short
range correlations (the diagonal marks the identity in time), and the long
range correlations corresponding to points distant from the main diagonal.
Besides the global impression given by the graphic appearance of the plot
(see Fig. 1 and Fig. 2, for the RQA plot of the protein and the fluid) the measures developed by Webber and Zbilut \cite{webber,trulla} allow for a
quantitative description of the recurrence structure of the plot. This is an important consideration, since visual inspection of plots can, at times, lend themselves to misinterpretation due to the vagaries of human perception. Additionally, quantification then allows for hypothesis testing.

The RQA descriptors are: REC = recurrence, which quantifies the percentage of the
plot occupied by recurrent points. It corresponds to the proportion of
recurrent pairs over all the possible pairs of epochs or, equivalently, the
proportion of pairwise distances below the chosen radius among all the
computed distances. DET = determinism, and is the percentage of recurrent
points that appear in sequence, forming diagonal line structures in the
distance matrix. DET corresponds to the amount of patches of recurrent
behavior in the studied series, i.e., to portions of the state space in
which the system resides for a time longer than expected by chance alone
(see \cite{zak,zbilut4}). This is a crucial point: a recurrence can, in
principle, be observed by chance whenever the system explores two nearby
points of its state space. On the contrary, the observation of recurrent
points consecutive in time (and then forming lines parallel to the main
diagonal) is an important signature of deterministic structuring \cite
{eckmann1,zbilut5}. The superposition between determinism and Lyapunov
exponents is a proof of this point \cite{eckmann1}. ENT = entropy, which is
defined in terms of the Shannon-Weaver formula for information entropy \cite
{webber,shannon} computed over the distribution of length of the lines of
recurrent points and measures the richness of deterministic structuring of
the series. LYAP is simply the length (in terms of consecutive points) of
the longest recurrent line in the plot. LYAP was found to accurately predict
({\it r }= 0.93) the value of the maximum Lyapunov exponent in a logistic
map going from a regular to chaotic regime \cite{trulla}. Finally, TREND is
the regression coefficient of the relation between time (in terms of
distance from the main diagonal) and the amount of recurrence. TREND
quantifies the fading away of recurrence going forward in time, and
represents a measure of stationarity \cite{webber}. Additionally, a time series of any of the RQA descriptors can be produced by windowing the originally embedded scalar series, and overlapping in a manner similar to time-varying spectral plots (see, e.g., \cite{trulla}, also Fig. 3,4,7).

\section{RESULTS}

All RQA descriptors were computed for both the LJ fluid and the protein MD
potential energy time series. In particular, for the fluid, data from a
simulation at $T=0.8$ (usual reduced unit) were used. To test for the null
hypothesis that the MD series are stochastic, the original trajectories were
randomly shuffled to obtain 30 copies of each series (Table I). The 95\%
confidence intervals for the RQA descriptors were computed, and the position
of the original series relative to the confidence intervals checked. Except
for REC (and here it is noted that the value for the MD simulation falls
within the range of obtained shufflings), the null hypothesis for LJ fluid
could not be rejected, pointing to the stochastic character of the fluid
simulation. For the protein MD, however, the RQA values were well beyond the
confidence limits of the shuffled series, thus demonstrating the presence of
strong ``phase information" for the protein dynamics. These features are qualitatively evident when looking at the recurrence plots of the protein (Fig. 1) and the LJ fluid (Fig. 2): while protein shows a very rich and intermingled texture, LJ plot is much more homogeneous.

\section{DISCUSSION}

A relevant portion of the theoretical work on MD was based upon LJ fluid
simulations performed by Rahman \cite{rahman}. These trajectories can be
defined as recurrent, Hamiltonian, mixing and K-flow, or following some
authors, ``Lyapunov unstable'' \cite{haile}. The simulated LJ system evolves
toward an equilibrium state, and the constant energy surface defined by the
initial conditions is accessible to the system itself. The motion of such a
system is at least mixing so as to sample all the explored surface.

In the case of the LJ fluid, the result obtained, in Eckmann, Kamphorst, and
Ruelle terms \cite{eckmann1}, can be defined as ``autonomous,'' i.e.,
typical of a system evolving following time independent equations: this
corresponds to our operational definition of a ``shuffling resistant''
potential energy time series. In fact, the RQA measures of the shuffled
series are not statistically different from the original series. This
behavior corresponds to a random-like sampling of the phase space of the
system, even if the sampling is driven by a deterministic ``engine'' such as
MD. These kinds of ``experiments'' were used by Verlet \cite{verlet,lebowitz}
to compute thermodynamic properties of fluids following a formalization
introduced by Birkhoff \cite{birkhoff,vonneumann}, and based on Boltzmann's
view of ergodicity assumptions.

The recurrence plot of the protein simulation (Fig. 1) immediately shows the
impossibility of direct averaging of the data. The simple visual inspection
of the plot highlights abrupt changes in the texture pointing to multiple
minima in the trajectory (rugged landscape as opposed to flat surface). This
point is underscored in Fig. 3, which depicts REC and DET in a windowed
series. REC presents as a rugged landscape, while DET displays several
discontinuities. More importantly, the shuffling procedure significantly
alters the numerical values of the RQA descriptors, thus demonstrating the
``shuffling sensitivity'' of the underlying trajectory. The ``ergodic''
constraints of complete accessibility and mixing are not sufficient to make
the system evolve to an equilibrium situation given the finite time of the
simulation, and as a result, the trajectory is trapped in a limited portion
of the energy surface. In such situations we can speak of metastable states 
\cite{ford,honeycutt}, which obviate the possibility of computing direct
averages. In order to compute physical measures on such simulations, the
local minima of the phase space must be revealed and their relative depth
estimated.

The thermalization algorithms used in MD are not guaranteed to preserve the
microcanonic properties of the system \cite{berendsen2}; nevertheless, we
think that the quantitative RQA measures can be correlated to the
thermodynamic properties of the system under investigation. In any case,
these measures allow us to derive some useful information about the shape
of the energy landscape of the simulation. As a matter of fact, the basic
algorithm of recurrence plots was developed by Eckmann and Ruelle \cite{eckmann2}
with the aim of reconstructing the dynamics relative to a time series in a
finite dimensional space, and of generating a tangent map of the
reconstructed dynamics in order to calculate Lyapunov exponents.

In the recurrence plot, a recurrence is scored (and the respective point
darkened) whenever $\left| x_j-x_i\right| <d$. We can think of this
inequality as the numerator of the incremental ratio, $dy/dx$, where $dx$
corresponds to the time interval between two sampled points in phase space.
Thus the recurrent points are corresponding segments of the trajectory going
through valleys of the multidimensional space on which the potential energy
is projected by the embedding procedure. These valleys have, by definition,
a low slope ($d$) (and thus they are recurrent in phase space). On a more
general note, the global texture of the plot is linked to the ruggedness of
the explored landscape. A dense texture is linked to smooth slopes and a
flat landscape, while a coarse texture points to steeper energetic barriers
with the lack of texture (no recurrent points at all) revealing transitions.
This qualitative picture is consistent with the analysis of the logistic map
by RQA \cite{trulla} where the phase transitions of the system (changes in
dynamical regime) were registered by the RQA measures.

Looking at the recurrence plot of the LJ fluid (Fig. 2) the loss of any
preferential directionality of the system (quantitatively proved by the
shuffling invariance) is clear. REC (Fig. 4) is considerably smoother
compared to Fig. 3, while DET is erratic, reflective of the very low
recurrence values. The directionality of Fig. 2 is highlighted when it is
compared with the plot relative to the annealing phase of a tripeptide (Fig.
5) (see \cite{giuliani}) where the directionality was imposed by a strong
order parameter; i.e., decrease of temperature. The protein recurrence plot
(Fig. 1) has a preferential directionality in time (shuffling sensitivity)
that allows us to appreciate the effective dimensionality of the explored
conformational space. With an adequate sampling time we can resolve both the
Frauenfelder substates \cite{nienhaus,ansari} in terms of large scale
typology of the plot (Figs. 6--7) and the features of the single substate in
terms of texture (Fig. 8).

It is important to note that, in practice, several different variables could have been chosen for the MD simulations, such as Coulomb energy, or van der Waals energy. Potential energy was chosen since it is a global descriptor of the physical motions that can be adequately sampled \cite{straub} relative to long relaxation times as compared to anharmonic motions \cite{steinbach}. Thus potential energy is suited for the studied time scale. In a related matter, while the generalized ergodic measure (GEM) of Straub and Thirumalai \cite{straub} can compute a distribution of energy barriers between substates, currently, this, is not possible with RQA. The advantage of RQA over GEM is its sensitivity to local fast transients, thus being much more suited to dynamically individuate phase transitions. In this sense, the two methods are complementary: GEM better for energy characterization, RQA better for time resolution.

Finally, it should be remarked that RQA allows for the identification of putatively important events along the studied dynamics using only one variable (potential energy) instead of many (e.g., single dihedral angles). The meaning of these events may, in fact, require specific analyses using variables such as vibrational times, ring flips of amino acids, folding times, protonation times, or diffusion times of water, and require a case by case approach. Our analysis is particularly useful in the calculation of free energy difference between reactant and product in which various MD simulations can reduce uncertainties in free energy calculations \cite{hodel}; although as with MD studies in general, it is still limited by the general time scale of physiologically important substates which are on the order of tens of micro to milli seconds.

In summary RQA seems to constitute a very promising tool for the
characterization of conformational substates in MD simulations.

\begin{figure}[tbp]
\caption{Protein recurrence plot.}
\label{Fig1}
\end{figure}

\begin{figure}[tbp]
\caption{LJ fluid recurrence plot.}
\label{Fig2}
\end{figure}

\begin{figure}[tbp]
\caption{Time series of protein MD simulation (top), with the respective REC
and DET plots, calculated on a 200 point windowed series overlapping one
point at a time.}
\label{Fig3}
\end{figure}

\begin{figure}[tbp]
\caption{Time series of LJ fluid simulation (top), with the respective REC
and DET plots, calculated as in Fig. 3.}
\label{Fig4}
\end{figure}

\begin{figure}[tbp]
\caption{Simulated annealing tripeptide recurrence plot.}
\label{Fig5}
\end{figure}

\begin{figure}[tbp]
\caption{Detail of Fig. 1 protein with two sections (squares, upper right
and lower left) clearly separated, representing two different microstates.}
\label{Fig6}
\end{figure}

\begin{figure}[tbp]
\caption{Section of protein MD simulation between microstates of Fig. 6
(top), with the respective DET plot. Note the clear divergence between the
two microstates at 78--81 ps.}
\label{Fig7}
\end{figure}

\begin{figure}[tbp]
\caption{Detail of lower left microstate of Fig. 6.}
\label{Fig8}
\end{figure}

\pagebreak

\begin{table}[tbp]
\caption{RQA results for shuffling.}
\label{tab1}
\begin{tabular}{lllll}
& Shuffled Mean & 95\% Conf. Int. & Range & MD Simulation \\ 
\multicolumn{5}{l}{LJ fluid} \\ 
\tableline REC & 0.76 & 0.74--0.78 & 0.66--0.87 & 0.688 \\ 
DET & 39.47 & 38.85--40.08 & 36.57--42.63 & 39.03 \\ 
ENT & 2.33 & 2.31--2.35 & 2.25--2.50 & 2.347 \\ 
LYAP & 17 & 16--18 & 14--21 & 16 \\ 
TREND & 0.009 & -0.03--0.05 & -0.23--0.24 & -0.005 \\ 
\multicolumn{5}{l}{Protein} \\ 
\tableline REC & 0.72 & 0.69--0.75 & 0.58--0.86 & 5.12 \\ 
DET & 39.20 & 38.08--40.31 & 33.20--44.88 & 69.48 \\ 
ENT & 2.31 & 2.27--2.36 & 2.06--2.62 & 3.25 \\ 
LYAP & 14 & 13--15 & 10--21 & 31 \\ 
TREND & 0.033 & -0.08--0.15 & -0.45--0.81 & -2.27
\end{tabular}
\end{table}

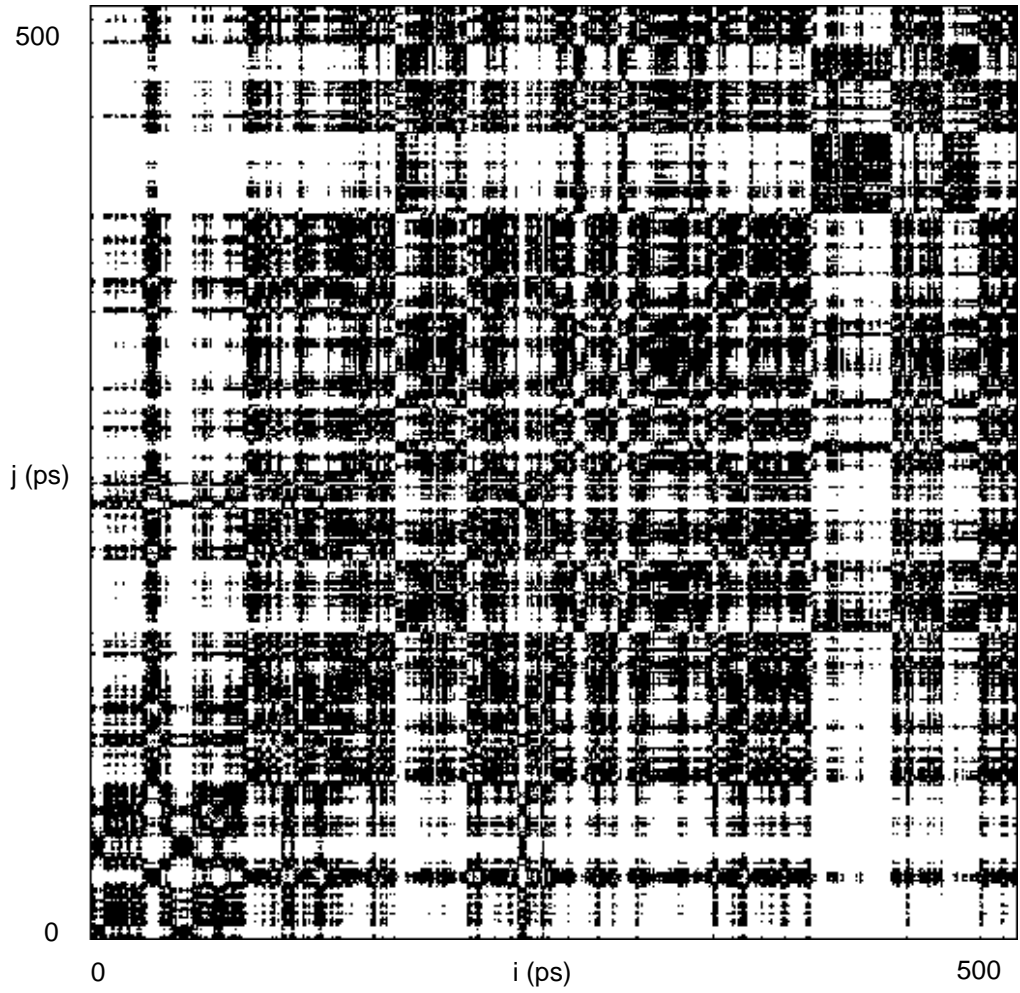
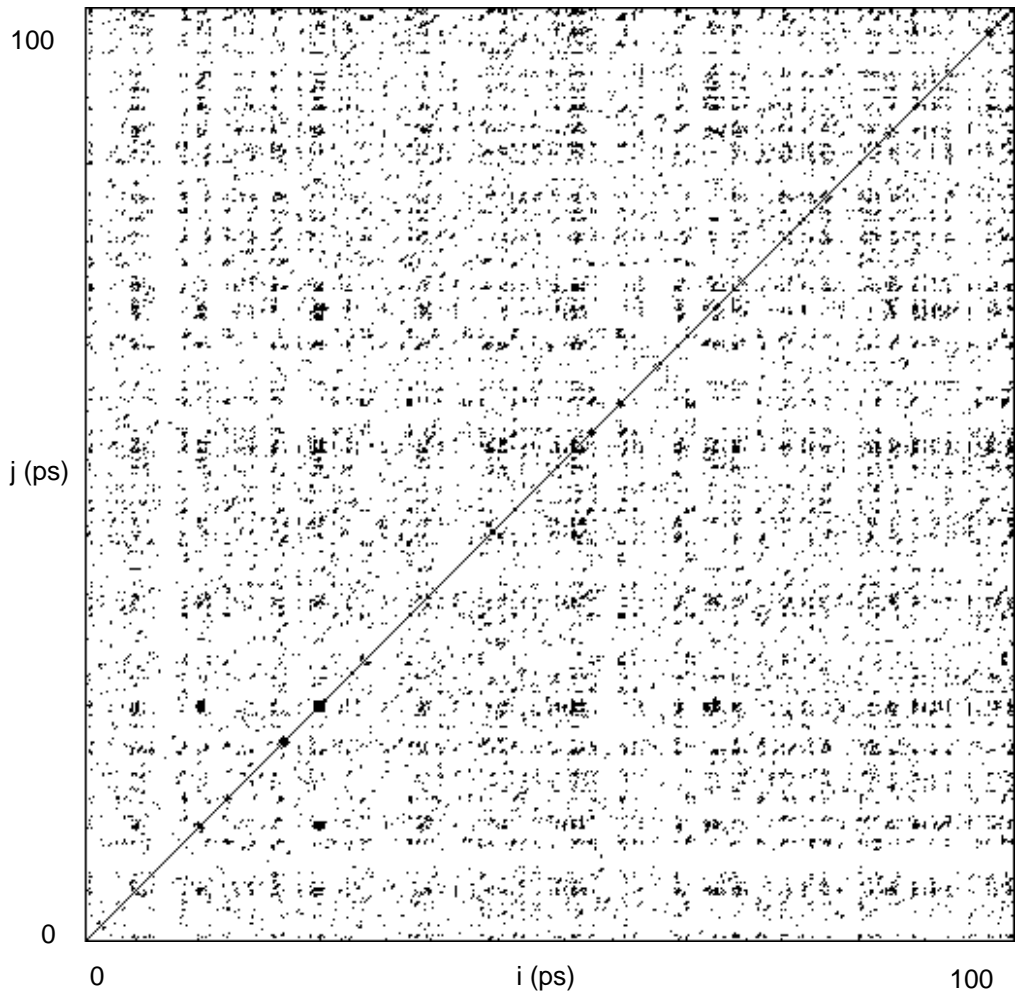
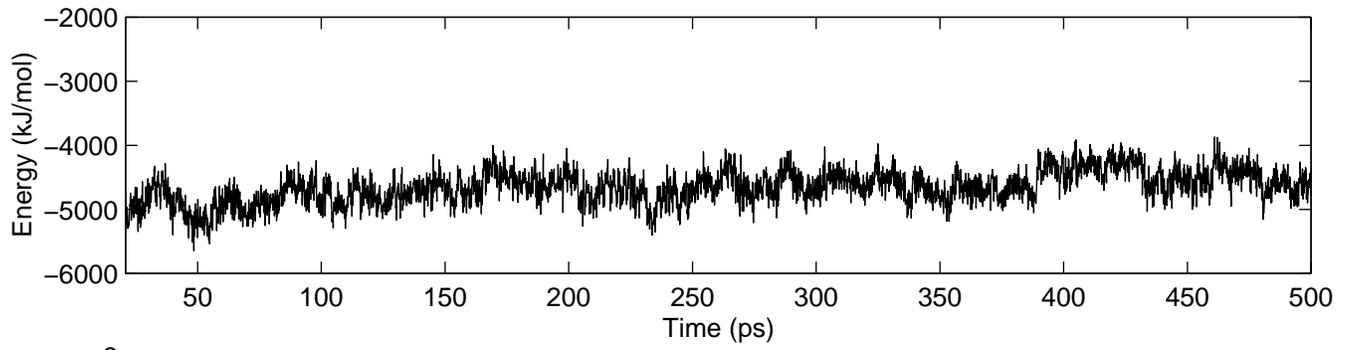
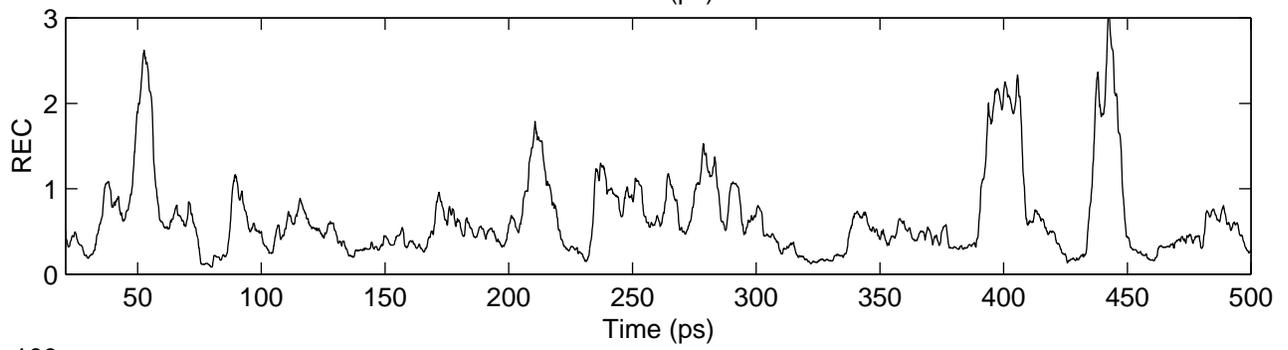
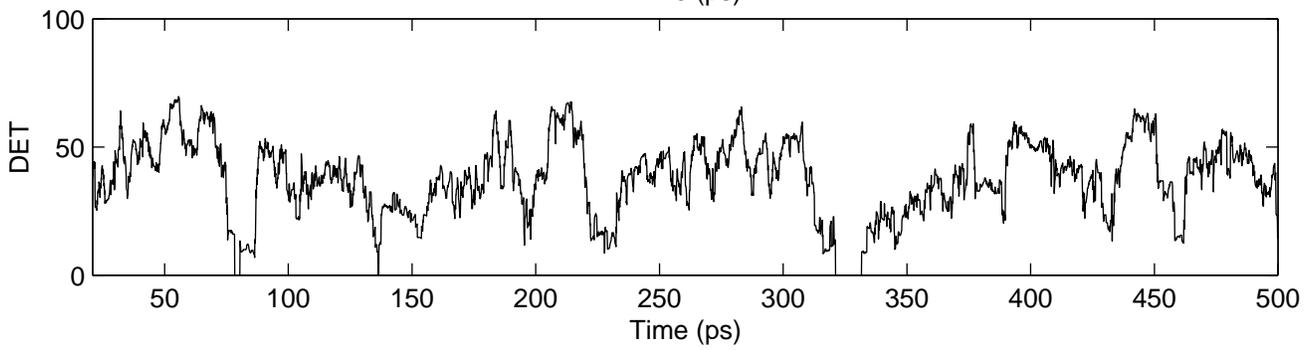
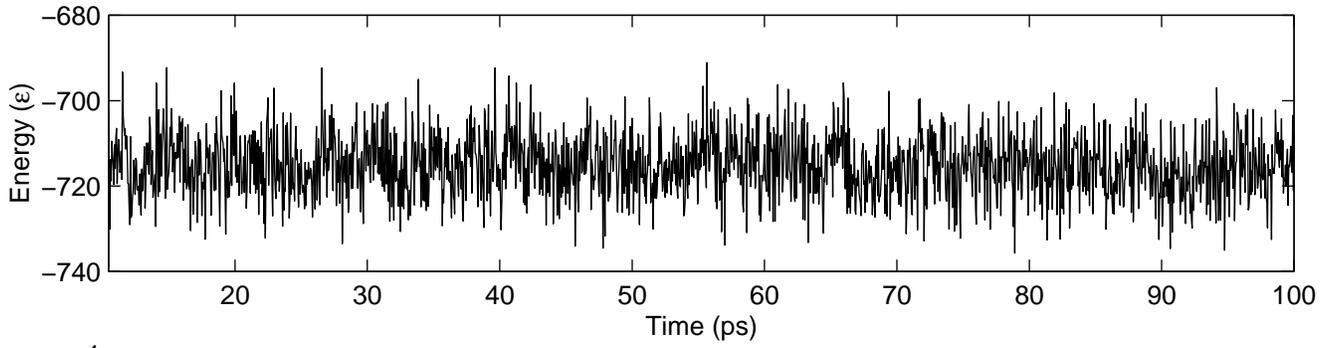
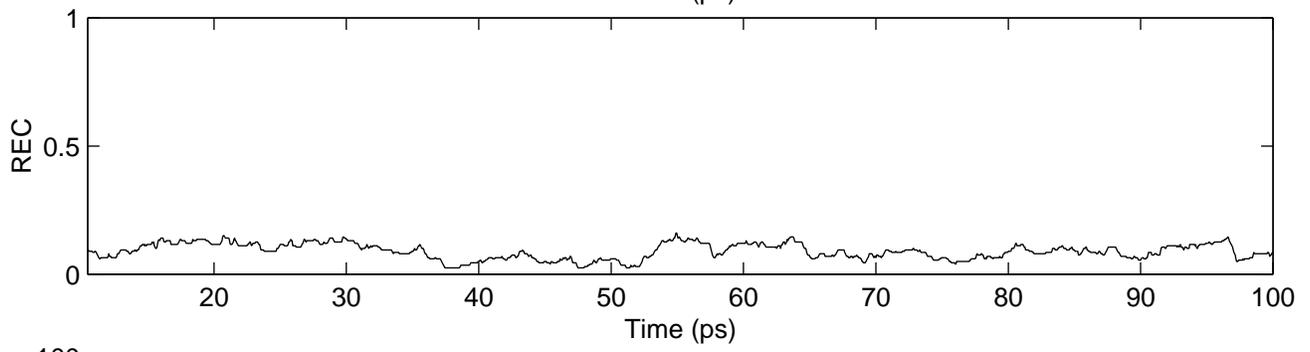
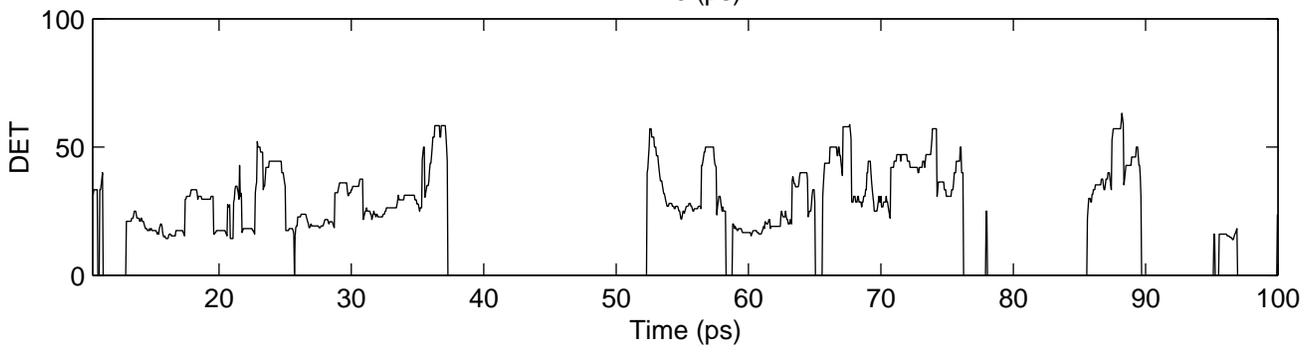
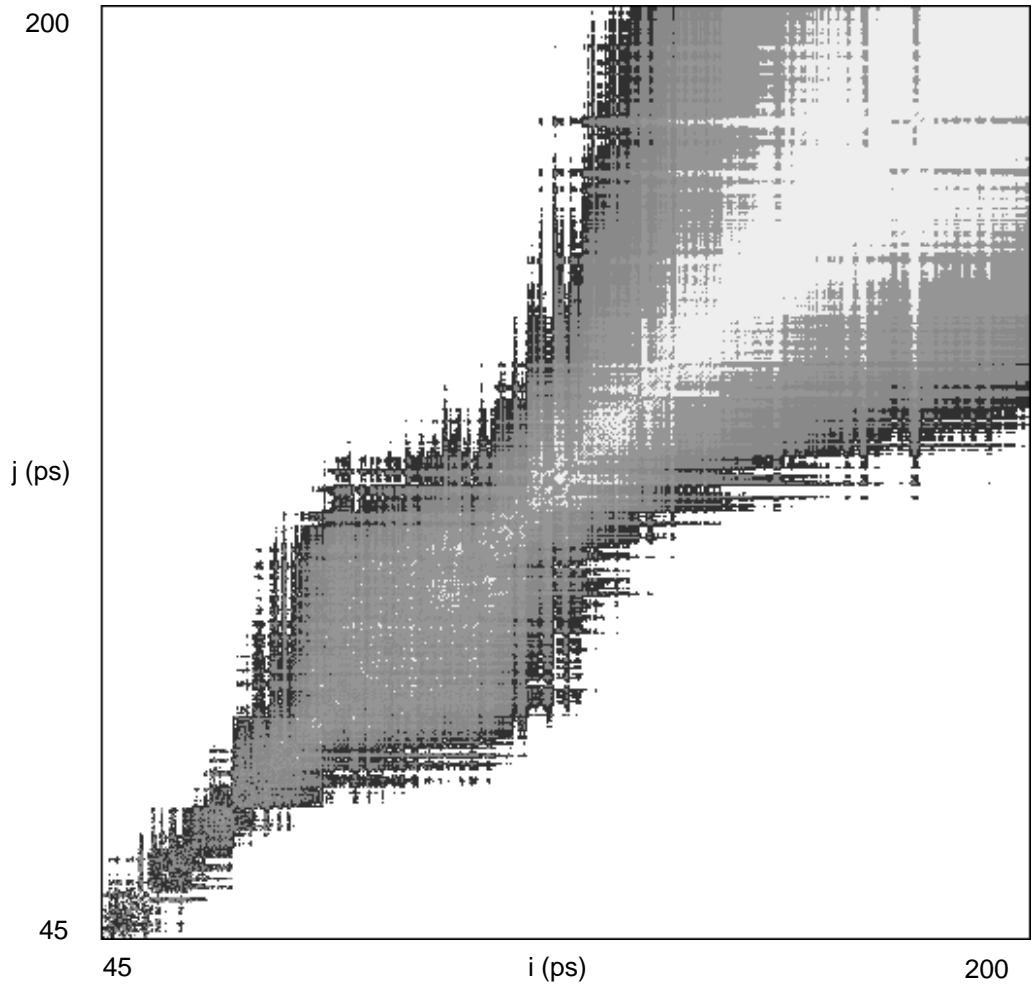
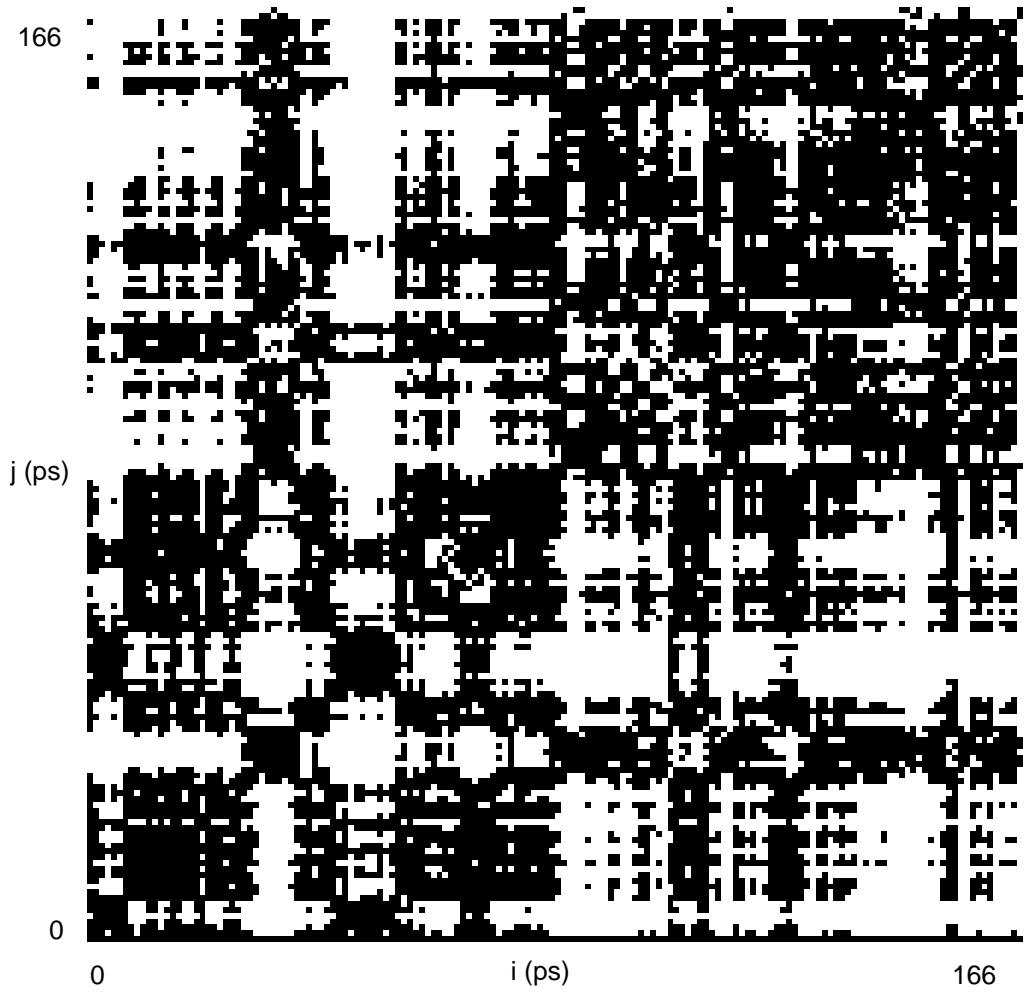
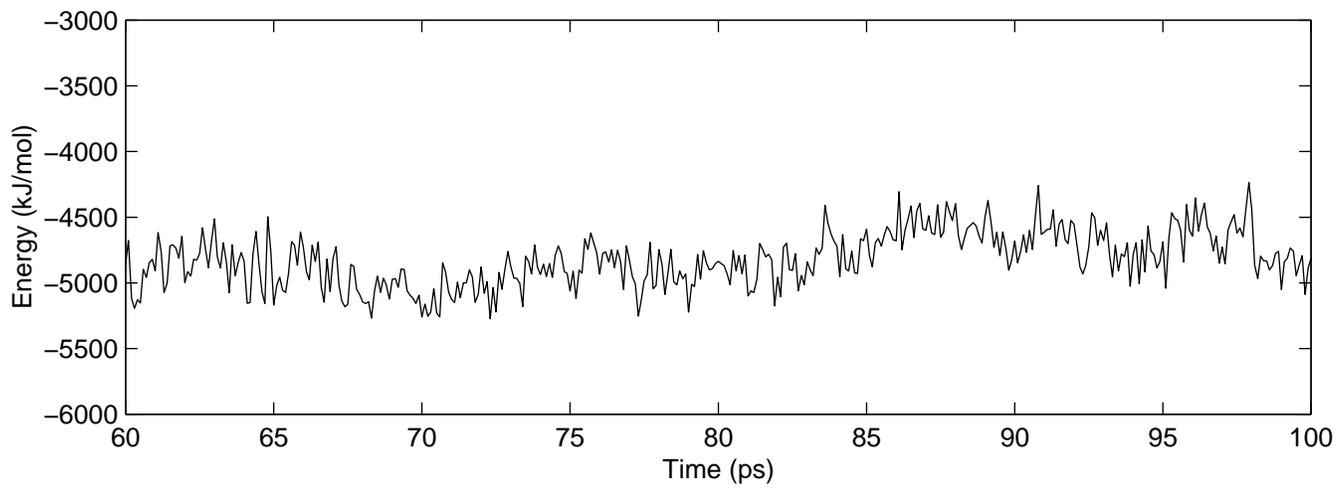
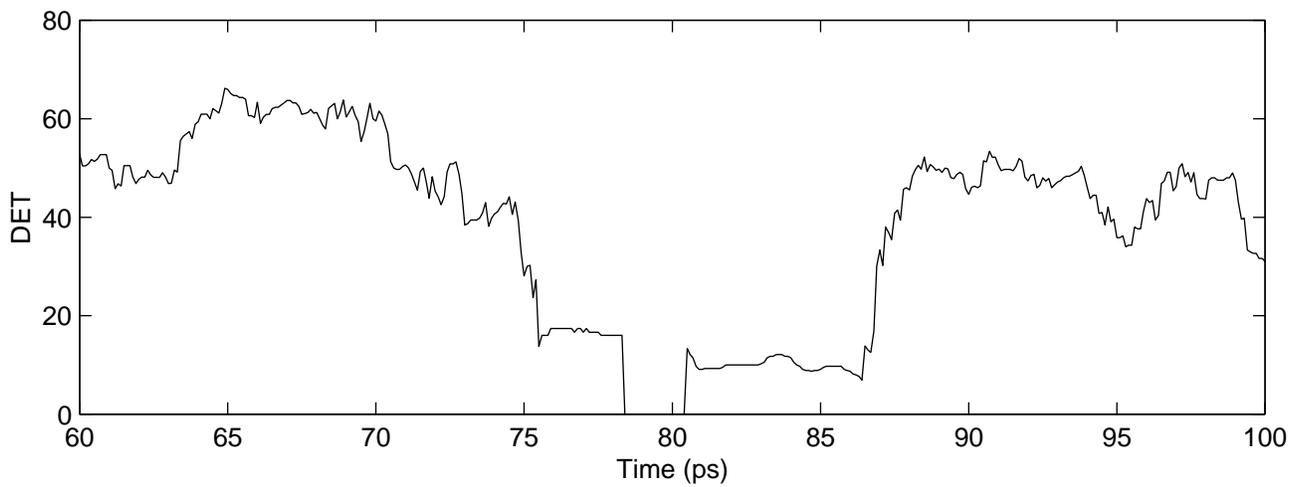
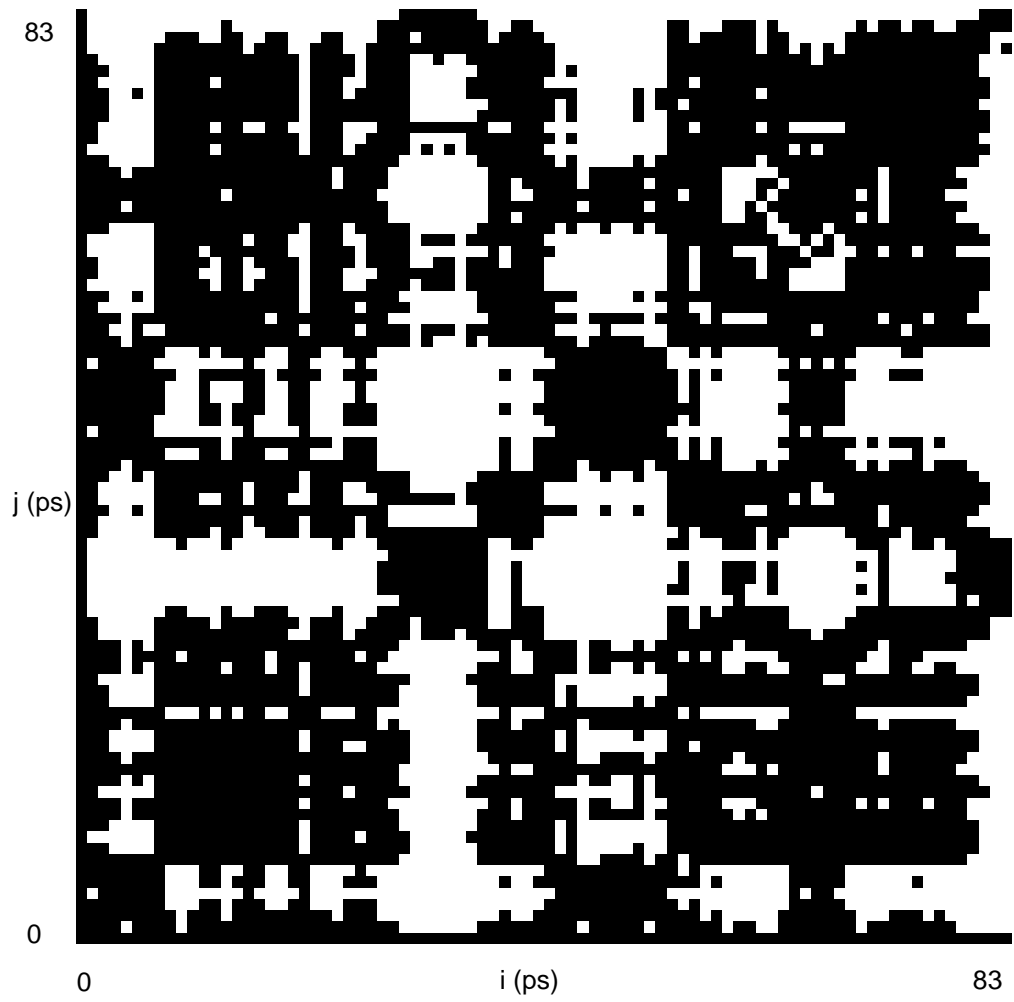
\end{document}